\documentclass[12pt]{article}
\begin{document}
\title{THE INTERFACE BETWEEN CLASSICAL ELECTRODYNAMICS AND QUANTUM THEORY}
\author{B.G. Sidharth\\
Centre for Applicable Mathematics \& Computer Sciences\\
B.M. Birla Science Centre, Adarsh Nagar, Hyderabad - 500 063 (India)}
\date{}
\maketitle
\begin{abstract}
In this paper we discuss in detail the interface between Classical Electrodynamics and Quantum Theory, which shows up as well known unphysical phenomena at the Compton scale in both the theories and argue that the photon of the electromagnetic field can be considered to be a composite of a massless Dirac particle and its anti particle.
\end{abstract}
\section{Introduction}
A long time ago, Darwin showed that the massless, force free Dirac theory was formally identical to source free electrodynamics in a vaccuum [1] In the absence of a suitable physical interpretation this mathematical identity has for long been considered to be a mere mathematical coincidence (Cf.ref.[1]). After all, photons are spin one particles, while the Dirac equation represents spin half particles. At the same time, it has also been recognized for a long time - Einstein and Meyer were one of the first to point this out - that the spinorial representation of the Lorentz group is more fundamental than the vectorial representation [2].

In the light of the above observations we would now like to point out that the above circumstance is not a mere coincidence, but has a definite physical interpretation.

We firstly make some preliminary remarks: Both in electromagnetic theory and in the Dirac theory, the D'Alembertian equation

\begin{equation}
D \psi_\mu = 0\label{e1}
\end{equation}

where $D$ is the D'Alembertian operator,
is satisfied by the respective components. This is merely an expression of Lorentz invariance. At this point the two theories diverge. This is because an equation like (\ref{e1}) requires the value of $\psi$ at say $t = 0$ and so also the value of $\frac{\partial \psi}{\partial t}$ for specifying the solution. This does not pose any problem in electromagnetic theory, but is not acceptable in Quantum theory, because the Quantum Mechanical wave function $\psi$ contains as complete a description of the state as is possible and there is no room for derivatives as initial conditions. This is also the reason why (\ref{e1}), or the Quantum Mechanical Klein Gordon equation gives negative probability densities. So the order of (\ref{e1}) needs to be depressed to make it a first order equation, which infact is the starting point of the Dirac theory and leads to the Dirac equation, [3],

\begin{equation}
\left(\gamma^\mu p_\mu - m\right) \psi = 0\label{e2}
\end{equation}

It may be mentioned that two component spinors belonging to the representation 
$$D^{(\frac{1}{2}0)} \mbox{or} D^{(0\frac{1}{2})}$$
of the Lorentz group are solutions of the Dirac equation (\ref{e2}). But these are no longer invariant under reflections [4]. It is to preserve this invariance that we have to consider the $4 \times 4$ representation
$$D^{(\frac{1}{2}0)} \oplus  D^{(0 \frac{1}{2})}$$
Under reflections, the two spinors transform into each other thus maintaining the overall invariance [5]. We also note that, as is known [6], the Maxwell equations can also be written in the form of neutrino equations. Defining a four vector such that

\begin{equation}
\chi_j = E_j + \imath B_j, \chi_0 = 0\label{e3}
\end{equation}

we can rewrite the Maxwell equations in the form

\begin{equation}
\beta_\mu \frac{\partial \chi_\nu}{\partial x_\mu} = - \frac{1}{c} j_\nu\label{e4}
\end{equation}

where in a particular representation, for example,

$$\beta_0 = I X I, \quad  \beta_1 = -\sigma_3 \otimes  \sigma_2,$$
$$\beta_2 = \sigma_2 \otimes I, \quad \beta_3 = \sigma_1 \otimes \sigma_2,$$
the $\sigma$'s being the Pauli matrices and wherein for our source free vaccuum case, the current four vector on the right hand side of equation (\ref{e4}) vanishes. It is easy to show that the four component equation (\ref{e4}) breaks down into two two component neutrino like equations, except that both these equations are coupled owing to the additional condition $\chi_0 = 0$ in (\ref{e3}). This has been the problem in identifying (\ref{e4}) with the Dirac theory.

\section{Photons}

In the above context let us now approach the above considerations from the opposite  point of view, that of the Dirac equation. It is well known that the four linearly independent four spinor Dirac wave functions are given by [7], apart from multiplicative factors,

\begin{equation}
\left[\begin{array}{ll}
1 \\ 0 \\ \frac{p_zc}{E + mc^2} \\ \frac{p+c}{E = mc^2}
\end{array}\right] 
\left[\begin{array}{ll}
0 \\ 1 \\ \frac{p-c}{E + mc^2} \\ \frac{-p_zc}{E + mc^2}
\end{array}\right]
\left[\begin{array}{ll}
\frac{p_z c}{E + mc^2} \\ \frac{p + c}{E + mc^2} \\ 1 \\ 0
\end{array}\right]
\left[\begin{array}{ll}
\frac{p+c}{E+mc^2} \\ \frac{-p_z c}{E + mc^2} \\ 0 \\ 1
\end{array}\right]\label{e5}
\end{equation}

where $p_z$ is the $z$ component of the momentum and

$$p_\pm = p_x \pm \imath p_y,$$

in a representation given by,

$$\gamma_\imath = \gamma_0 \left[\begin{array}{ll}0 \quad \sigma_\imath \\
\sigma_\imath \quad 0
\end{array}\right], \gamma_0 = \left[\begin{array}{ll}
1 \quad 0 \\
0 \quad -1
\end{array}\right]$$

the $\sigma$'s being the Pauli matrices.

If we consider the $z$ axis to be in the direction of motion, for simplicity and take the limit $m \to 0$ in (\ref{e5}), the spinors in (\ref{e5}) become,

\begin{equation}
\psi_1 = \left[\begin{array}{ll}
1 \\ 0 \\ 1\\ 0
\end{array}\right]
\psi_2 = \left[\begin{array}{ll}
0 \\ 1 \\ 0 \\ -1
\end{array}\right]
\psi_3 = \left[\begin{array}{ll}
1 \\ 0 \\ 1 \\ 0
\end{array}\right]
\psi_4 = \left[\begin{array}{ll}
0 \\ -1 \\ 0 \\ 1
\end{array}\right]
\label{e6}
\end{equation}

It should be noticed that in (\ref{e6}) surprisingly  $\psi_1 = \psi_3$, so that effectively, one of the spinors vanishes exactly as in the case of the solutions $\chi$ of (\ref{e4}).(The mass zero four component Dirac spinor does not represent a neutrino unless an auxiliary condition, which effectively destroys the lower two or upper two components is imposed [5]). On the other hand, without such an auxiliary condition, it can now be seen from the above considerations that the source free vaccuum electromagnetic field can be considered to be a composite of a neutrino and an anti neutrino. It may be mentioned that the possibility of Bosons being bound states of Fermions, rather than primary has been discussed by the author and other scholars [8].\\
We now push the above considerations further and try to understand the connection of electrodynamics with spinorial representations, from a different point of view.

Let us start with the Lorentz Dirac equation of Classical Electrodynamics [9]

\begin{equation}
ma^\mu = F^\mu_{in} + F^\mu_{ext} + \Gamma^\mu\label{e7}
\end{equation}

where
$$F^\mu_{in} = \frac{\epsilon}{c} F^{\mu \nu}_{in} v_\nu$$
and $\Gamma^\mu$ is the well known  Abraham radiation reaction four vector.

It must be mentioned that this is the relativistic version of Lorentz's earlier equation. The Abraham radiation reaction four vector represents the energy loss by radiation. It is interesting that the mass $m$ in (\ref{e7}) consists of the original electromagnetic mass of Lorentz which tends to infinity as the size of the electron tends to a point, but this infinity is absorbed into a neutral mass which is also present in this term. As in later renormalization theory, this is the case where we preserve the particle as a point concept by absorbing infinities suitably.

Unfortunately (\ref{e7}) displays a number of difficulties: Runaway solutions, which are intimately connected to the point electron concept and the breakdown in causality. However it has been shown in detail that these difficulties, including the non locality are phenomena which can be meaningfully interpreted within the context of the Feynman-Wheeler Action at a Distance Theory (Cf.refs.[9,10,11,12,13] for details).

It is at this stage that Classical Electrodynamics forms an interface with Quantum Theory: In Quantum Theory there is exactly this non locality or breakdown of causality at the Compton scale [14,15]. As discussed in detail (Cf.ref.[15]) this is indicative of the limits of our observation at the Compton scale, and once there is such a limiting scale of measurement, as is well known (Cf.[16,17], we are lead to a Lorentz invariant non commutative geometry

\begin{equation}
[x,y] = 0(l^2), [x,p_x] = \imath \hbar [1 +(l/\hbar )^2 p^2_x | etc.,\label{e8}
\end{equation}

where $l$ is the Compton wavelength. Infact the original motivation for investigating (\ref{e8}) was the elimination of the infinities which we saw above. Incidentally the noncommutativity (\ref{e8}), whlich is otherwise classical, already leads to the Quantum theory, even in the limit $l \to 0$.

It is very interesting that given (\ref{e8}), we can directly deduce the Dirac equation [18,15].  In other words (\ref{e8}), which we can view upon as a junction equation between Classical Electrodynamics and Quantum Theory, leads to the bi-spinorial formulation discussed earlier. Starting from the Dirac equation (\ref{e2}), we can infact recover electromagnetism. This has been discussed in detail [19,20], but to sum up, the Dirac bi-spinor $\left(\begin{array}{ll}\Theta \\ \chi
\end{array}\right)$ has the so called positive energy components $\Theta$ and negative energy components $\chi$. It is also known that while the $\Theta$ components dominate well outside the Compton scale, it is the $\chi$ components that predominate at and near Compton scales. Moreover under reflections, it is known that [21]

$$\chi \to -\chi,$$

so that we have

\begin{equation}
\frac{\partial \chi}{\partial x^\mu} \to \frac{\imath}{\hbar} \left[\frac{h}{\imath} \frac{\partial}{\partial x^\mu} + \imath N A^\mu\right]\chi\label{e9}
\end{equation}

(\ref{e9}) can be shown to lead to the electromagnetic potential

\begin{equation}
A^\mu = h \Gamma_\sigma^{\mu \sigma} = \hbar \frac{\partial}{\partial x^\mu} log(\sqrt{|g|})\label{e10}
\end{equation}

which is identical to the electromagnetic potential of the Weyl theory [22,23].

At this point it may be mentioned that the original Weyl Theory, though formally correct was not accepted because it appeared ad hoc. Moreover this theory was set in the usual classical space time. We on the other hand have a different starting point, namely the Compton scale non commutativity, (\ref{e8}) which is also tied up with the interpretation of the Lorentz-Dirac equation of classical electrodynamics. Let us put this in perspective (Cf.ref.[24] for details): Given (\ref{e8}), an infinitesimal Lorentz transformation of the wave function, (\ref{e8}) leads to 

\begin{equation}
\psi ' (x_j) = \left[ 1 + \imath \epsilon \left( \imath x_k \frac{\partial}{\partial x_k}\right) + 0(\epsilon^2)\right] \psi (x_j)\label{e11}
\end{equation}

On the one hand, at the Compton scale $l$, (\ref{e11}) leads to the Dirac equation as already mentioned. On the other hand, we could view (\ref{e11}) as a transformation of coordinates

\begin{equation}
x^\mu \to \gamma^{(\mu)} x^{(\mu)}\label{e12}
\end{equation}

where the brackets with the superscripts have been shown to indicate the fact taht there is no summation over repeated indices. The equation (\ref{e12}) was studied a long time ago by Bade and Jehle [25]. This infact leads to the familiar Clifford Algebra with a representation for the $\gamma$s being

\begin{equation}
\gamma^k = \sqrt{2} \left(\begin{array}{ll}
0 \quad \sigma^k \\
\sigma^{k*} \quad 0
\end{array}\right)\label{e13}
\end{equation}

the $\sigma$s being the Pauli matrices. As discussed by Bade and Jehle we could take the $\sigma$s or $\gamma$s in (\ref{e13}) as the components of a contravariant world vector, or equivalently we could consider them to be fixed matrices and for covariance attribute new transformation properties to the wave function, which thereby becomes the usual Dirac bi-spinor. The latter point of view has been considered traditionally, with, space time coordinates retaining their classical character. On the other hand, if equivalently, we consider the former alternative, then we return to the non commutative geometry (\ref{e8}). So, either we have the usual picture in which the Dirac spinor conceals the non commutative character (\ref{e8}) leading to equations like (\ref{e9}) or (\ref{e10}) reminiscent of the Weyl geometry, or equivalently we consider the non commutativity (\ref{e8}), which as seen above can also arise from the classical Lorentz theory of electrodynamics.

Interestingly (\ref{e10}) can also be deduced directly from (\ref{e8}) without having to go via the Dirac theory (Cf.[19]).

Finally we can get an insight into the spinorial feature from a different angle. Zakrzewski [26] has shown that for a classical spinning particle we have the Poisson bracket relation

$$\{x^j,x^k\} = \frac{1}{m^2} R^{jk}, (c = 1),$$

where $R^{jk}$ is the spin. The usual passage from Poisson brackets to Quantum commutators leads us back to (\ref{e8})([11,27]).

In summary, we have shown the interface between Classical Electrodynamics and Quantum Theory on the one hand and have argued that the photon of the electromagnetic field can be considered to be a composite of a massless Dirac particle and its anti particle. For a justification including the decay mode, Cf.refs. [28,29].

{\large \bf{Addendum}}: An equation like (\ref{e8}) expressing non-commutativity of coordinates or differentials can be viewed in the following manner. The metric tensor breaks up into two parts, $g_{\mu \nu}$ and $g'_{\mu \nu}$ while $g'_{\mu \nu}$ is of the conventional kind, $g_{\mu \nu}$ can be written as 
$h^{\delta \sigma} \epsilon_{\delta \sigma \mu \nu}$, where $h^{\delta \sigma}$ is an antisymmetric tensor and $\epsilon$ is the Levi-Civita tensor denslity. This is immediately recognized as Weyl's guage geometry formulation.

\end{document}